\documentclass[twoside]{dis09}
\usepackage[latin1]{inputenc}
\usepackage[dvips]{graphicx,epsfig,color}
\usepackage{wrapfig,rotating}
\usepackage{amssymb,amsmath,array}
\def\Journal#1#2#3#4{{#1} {\bf #2} (#4) #3}

\def\PRD{{Phys. Rev.}    {\bf D}}

\newcommand {\lapprox}
   {\raisebox{-0.7ex}{$\stackrel {\textstyle<}{\sim}$}}

\begin{document}
\title{Scattering of Heavy Stable Exotic Hadrons}

\author{David Milstead
%
%
\vspace{.3cm}\\
%
Fysikum, Stockholms Universitet \\
10691 Stockholm, Sweden.
%
}

\maketitle

\begin{abstract}
Exotic stable massive particles (SMP) are proposed in a number of scenarios of physics beyond the Standard Model.
 It is important that LHC experiments are able to detect hadronic SMP with masses around the TeV scale. To do this,
  an understanding of the interactions
   of SMPs in matter is required. In this paper a Regge-based model of $R$-hadron scattering is
   extended and implemented in {\sc Geant-4}.
\end{abstract}

\begin{center}
{\it Prepared for Proceedings of 17$^{th}$ International Workshop on Deep-Inelastic Scattering and Related Subjects, Madrid, 2009.}
\end{center}

\section{Introduction}
The observation of exotic stable\footnote{The term stable is taken to
mean that the particle will not decay during its traversal of a
detector.} massive particles (SMPs) with electric, magnetic or colour charge (or combinations thereof) would be of fundamental significance. Searches are therefore routinely made for SMPs bound in matter, and for SMPs at cosmic ray and collider experiments~\cite{Amsler:2008zzb,Drees:1990yw,Perl:2001xi,Fairbairn:2006gg}.
Furthermore, SMPs with masses at the TeV scale are anticipated in a number of beyond-the-Standard-Model scenarios, such as supersymmetry and universal extra dimensions~\cite{Fairbairn:2006gg}. SMP searches are thus a key component of the early data exploitation program of the LHC experiments~\cite{Aad:2009wy}. To facilitate such searches, models are needed of the interactions of SMP as they pass through matter, since such interactions can give rise to a range of different event topologies~\cite{Fairbairn:2006gg}. Although robust phenomenological approaches are available to predict the scattering of electromagnetically charged SMPs, it is not clear how the interactions of hadronic SMPs (hereafter termed $R$-hadrons\footnote{The term $R$-hadron has its origin in the $R$-parity quantum number in supersymmetry theories and the work in this paper will be presented in the context of supersymmetric particles. However, the results are generally applicable to stable heavy exotic hadrons.}) should
 be treated owing to uncertainties associated with the strong scattering processes and the hadronic mass hierarchy. This paper describes an implementation within {\sc Geant-4}~\cite{Agostinelli:2002hh} and extension of a Regge-based model~\cite{deBoer:2007ii} for hadronic scattering. The paper is a summary of work contained in Ref.~\cite{Mackeprang:2009ad}.

To aid the development of search strategies, it is important that experiments have access to Monte Carlo models which span as fully as possible the
 range of conceivable signatures which could be associated with $R$-hadron production. At present, one model is available within {\sc Geant-4}~\cite{Mackeprang:2006gx}. This model, hereafter termed the \emph{generic model}, is based on a black disk approximation~\cite{Kraan:2004tz} and employs pragmatic assumptions regarding possible scattering processes and
 stable $R$-hadron species. This work concerns an approach, hereafter termed the \emph{Regge model}, which uses triple regge formalism
  to predict the scattering of $R$-hadrons formed from exotic colour triplet squarks (stop-like ($\tilde{t}$) and sbottom-like ($\tilde{b}$)) and electrically neutral gluino-like ($\tilde{g}$) colour octet particles.
Furthermore, the model uses a different set of well motivated assumptions regarding $R$-hadron mass hierarchies  than those employed in
the generic model.

\section{Heavy hadron mass hierarchy}\label{sec:massh}
Although there exist no complete calculations of expected $R$-hadron mass spectra, estimates have been
made using a variety of approaches such as the bag model~\cite{Chanowitz:1983ci,Farrar:1984gk,Buccella:1985cs} and
lattice QCD~\cite{Foster:1998wu}.
These predictions, together with with measurements of heavy hadron masses, make it possible to approximately determine
those features of the mass hierarchies which are
most relevant for the modelling of $R$-hadron scattering in material. Of particular interest are the masses of
the lowest lying states, to which higher mass
particles would be expected to decay before being able to interact hadronically.

Given the observed spin and flavour independence of interactions of heavy quarks with light
 quarks~\cite{Isgur:1989vq} the lightest meson and baryon states can be inferred~\cite{Gates:1999ei}  from the measured mass spectra
 of charm and bottom hadrons. Thus, charged and neutral squark-based mesons would be approximately mass degenerate and the lightest baryon
state to which the other baryons would decay would be the scalar singlet $\tilde{q}ud$ i.e. the exotic
equivalent of the $\Lambda_c$,$\Lambda_b$ baryons. This picture is supported by calculations of hyperfine splittings of
$R$-hadron masses~\cite{Gates:1999ei,Kraan:2004tz}.

The basic properties of gluino $R$-hadrons may not be similarly inferred from Standard Model hadronic mass spectra and a greater
reliance on phenomenological approaches is required to estimate the stable states. Possible hadrons include gluino $R$-mesons \\
 ($\tilde{g}u\bar{u}^0,\tilde{g}u\bar{d}^0,\tilde{g}d\bar{d}^0$) and  gluino-gluon
states ($\tilde{g}g$). Bag model~\cite{Chanowitz:1983ci} and lattice QCD~\cite{Foster:1998wu} calculations estimate that
the masses of lowest lying states of each of these hadrons differ by less than a pion mass. Detailed calculations of the lowest
lying baryon states~\cite{Farrar:1984gk,Buccella:1985cs} using the bag model predict that the lightest state is the singlet
 $\tilde{g}uds$. Consequently, higher mass states would decay weakly into the singlet state over a time scale of $\sim 10^{-10}$s~\cite{Buccella:1985cs}.

For the work presented here, we use the squark and gluino $R$-hadron mass hierarchy assumptions given above and use a simplified
mass spectra which comprises only those  $R$-hadrons deemed to be stable.
This leads to $R$-hadron event topologies which would be challenging for the one of the most common collider search methods, which looks
for a slow muon-like object~\cite{Aad:2009wy,Acosta:2002ju,Abazov:2008qu,Aaltonen:2009ke,Mermod:2009ct}. This is due to the $R$-baryon production processes
 which take place as $R$-hadron pass through matter. Such
processes, which, as discussed in the next section, imply that a sbottom or gluino $R$-hadron, irrespective of its
state as it entered a calorimeter, is likely to leave as an uncharged object and therefore escape detection in an outer muon chamber.

\subsection{Regge Model}
Since the central picture of a low energy light quark system interacting with a stationary nucleon~\cite{Kraan:2004tz}, $R$-hadron scattering
can be treated with the phenomenology used to describe low energy hadron-hadron scattering data~\cite{deBoer:2007ii,Baer,mafi}, as
is done in the Regge model of which more details can be found in Ref.~\cite{deBoer:2007ii}.  The Regge model was originally developed to describe squark $R$-hadron scattering though has
been extended here to also treat gluino $R$-hadrons. This model assumes the stable states described in the previous section.

\begin{figure}
\centerline{\includegraphics[width=0.85\columnwidth]{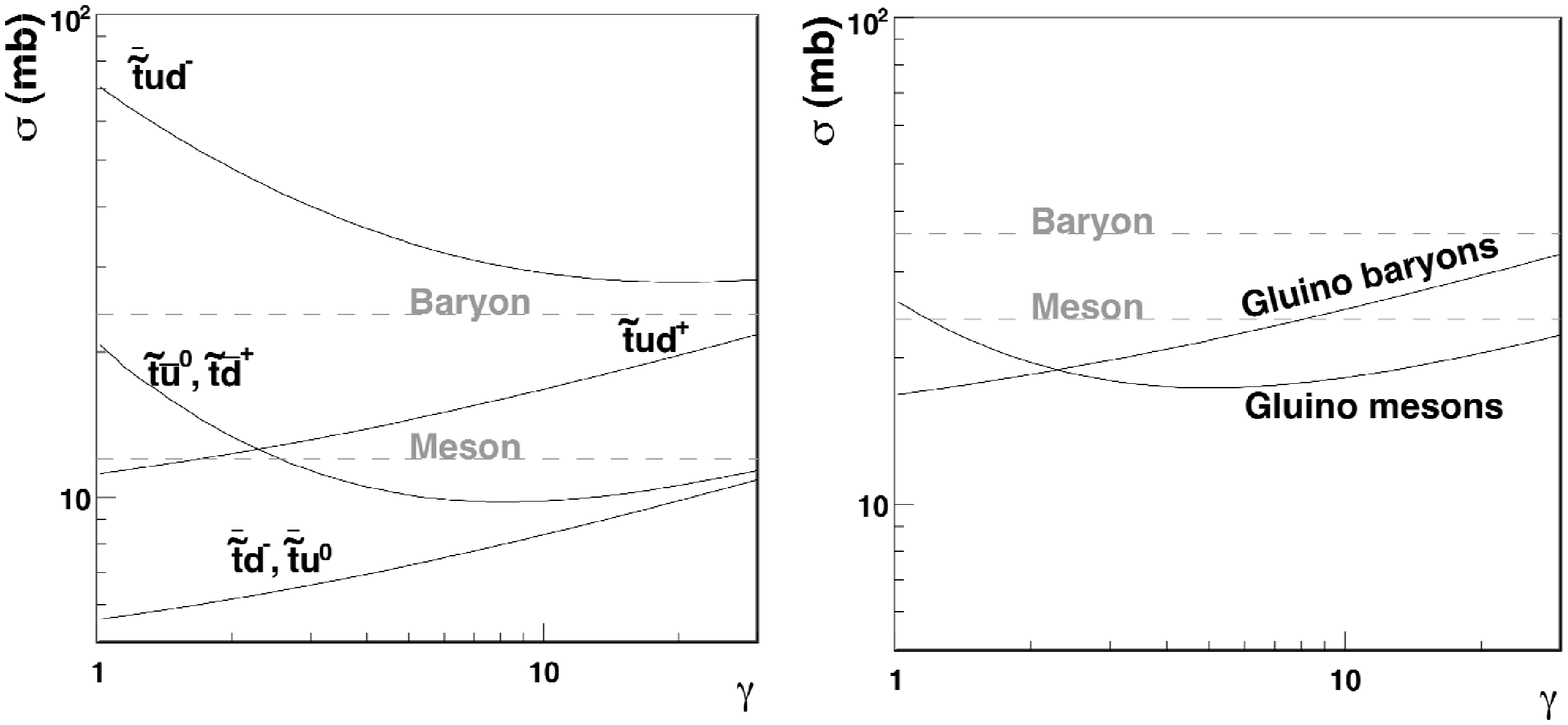}}
\vspace{-1.6cm}
\caption{Cross sections for stop-based and gluino-based $R$-hadrons. The predictions are shown for the Regge model (solid lines) and the generic model (dashed lines).}\label{Fig:fig1}
\end{figure}

Using parameters fitted to low energy hadron-hadron data, the Regge model makes predictions for $R$-hadron scattering cross sections,
together with energy loss calculations based on the triple regge formalism.
Fig.~\ref{Fig:fig1} (left) shows the the model predictions of the scattering cross sections of a squark-based $R$-hadron off
a stationary nucleon within a
nucleus comprising equal numbers of neutrons and protons. The cross section is shown for different types of squark-based
$R$-hadrons as a function of the Lorentz factor $\gamma$. As can be seen, there is a large cross section for antibaryon
($\bar{\tilde{q}}\bar{u}\bar{d}$) interactions which is due to a dominant annihilation process with a nucleon in the target.

It is also seen that, at lower values of $\gamma$ ($\gamma \lapprox 10$), the scattering cross section of
squark-based $R$-hadrons containing a light valence antiquark
($\tilde{q}\bar{u}^0,\tilde{q}\bar{d}$) is larger than for antisquark-based $R$-mesons, which arises from
the presence of reggeon exchange processes which are only permitted for $R$-hadrons containing a light anti-quark. Owing to the
presence of an additional light quark the scattering cross section for $R$-baryons (${\tilde{q}}{u}{d}$) is twice as large as
for the mesons with light quarks.

Upon an interaction, the probability that a  $R$-meson to $R$-meson process gives rise to charge
exchange is 50\%.  Similarly, the  probability of an interaction in which a $R$-meson becomes a
baryon is 10\%. Once a $R$-hadron becomes a baryon it stays in this state. Another process which must be taken into account is the oscillation of a neutral mesonic squark-based $R$-hadron
 into its antiparticle. Since the conversion rate would be model dependent we allow two possibilities here: zero mixing,
  in which no oscillations take place, and a maximal mixing scenario in which there is a 50\% probability that
  any neutral mesonic squark-based $R$-hadron which was produced would automatically be converted to its anti-particle.

In Fig.~\ref{Fig:fig1} (right) the predicted cross sections for gluino $R$-hadron species are shown. The baryon cross
section is 50\% greater than that of stop $R$-hadrons owing to the extra light quark in
the gluino $R$-hadron. Also, since the gluino meson contains a light quark and anti-quark,
then both regge and pomeron exchange processes are available. Again, a baryon number transfer probability of 10\% is used.
A probability of $\frac{1}{3}$ is assigned to $R$-meson to $R$-meson processes involving
charge exchanges of 0, $\pm e$ or $\pm 2e$.

\subsection{Implementation of Regge Model in {\sc Geant-4}}
In the dynamical picture of $R$-hadron scattering used by {\sc Geant}, the light quark system is decoupled
from the heavy spectator parton before interacting with a nucleon according  to {\sc Geant}'s  parameterised model for light hadrons.
In this way  secondary particles and so-called black tracks were generated. Following the interaction, the light quark system is
recombined with the heavy parton.

To implement the Regge model in {\sc Geant-4}, the software architecture used for the generic model was adapted.
The scattering cross sections and available baryon states were altered and the relative rates
of the available processes were adapted
to match the prescription in the above section. Using single particle simulations, it was verified that the main
 features (energy loss and charge exchange) of the Regge model were well reproduced in
 the {\sc Geant-4} implementation.

\section{Interactions and Energy Loss}\label{sec:res}
To estimate the effects of scattering on $R$-hadrons at the LHC simulations were made of the passage through
 iron of $R$-hadrons of 300 GeV mass. The kinematic distributions of the $R$-hadrons were given by the
  {\sc Pythia} generator which simulated the direct pair production of $\tilde{t}\bar{\tilde{t}}$, $\tilde{b}\bar{\tilde{b}}$ and $\tilde{g}{\tilde{g}}$ in proton-proton collisions at 14 TeV centre-of-mass energy.

\begin{figure}
\centerline{\includegraphics[width=0.55\columnwidth]{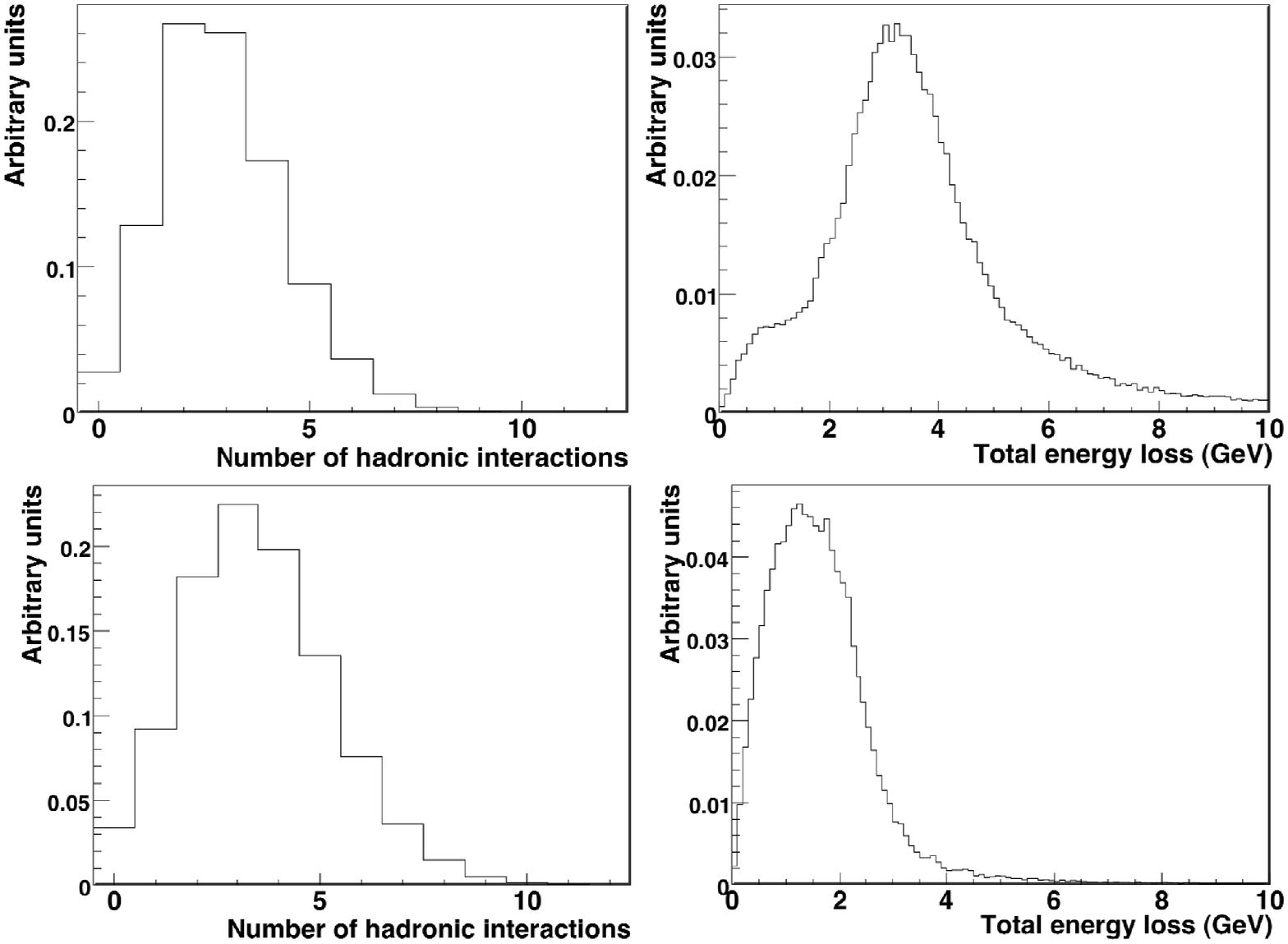}}
\caption{Number of hadronic interactions (left) and total energy loss (right) for
$R$-hadrons of 300 GeV mass sent through 1m of iron. The top (bottom) plots show predictions for stop (gluino) $R$-hadrons which entered the iron in the state $\tilde{t}\bar{d}$ ($\tilde{g}u\bar{d}$).}\label{Fig:fig2}
\end{figure}

The total energy loss for $R$-hadrons passing through 1m of iron is shown in Fig.~\ref{Fig:fig2}. Stop $R$-hadrons which start in the state $\tilde{t}\bar{d}$ typically lose more energy than gluino $R$-hadrons ($\tilde{g}u\bar{d}$) which start with the same charge. This occurs due to the rapid conversion of mesons to baryons. Since the gluino (stop) baryon is predicted to be of zero (positive) charge, it loses far less energy. Also shown is the number of hadronic interactions. Owing to the additional light quarks in the gluino $R$-hadrons more scattering occurs in this case than for the stop $R$-hadrons. A no-mixing scenario is assumed here.

\section{Summary}
An implementation and extension of a Regge-based approach to squark and gluino-based $R$-hadron scattering has been implemented in {\sc Geant-4}. The work allows the modelling of the energy loss and the transformation of $R$-hadrons as they propagate through material. Slides containing the presented work can be found in~\cite{url}

\section{Acknowledgments}
David Milstead is a Royal Swedish Academy Research Fellow supported by a grant
from the Knut and Alice Wallenberg Foundation.



\begin{footnotesize}



%

\end{footnotesize}


\end{document}